\documentclass[12pt,letterpaper]{article}  

\usepackage[includeheadfoot,
            marginratio={1:1,2:3}, 
            width=412pt, 
            height=688pt,]{geometry}

\usepackage{amsmath}
\usepackage{amsfonts}
\usepackage{amssymb}
\usepackage{mathtools}
\usepackage{mathalfa}
\usepackage{graphicx,caption,subfigure}
\usepackage{tikz} 
\usetikzlibrary{decorations.pathreplacing,calc}
\usepackage{booktabs}
\usepackage{adjustbox}
\usepackage[utf8]{inputenc}
 \usepackage[splitrule]{footmisc}

\usepackage{empheq}
\usepackage{paralist}
\usepackage{cite}
\usepackage[normalem]{ulem}

\usepackage[colorlinks=false,urlbordercolor=red]{hyperref}
\usepackage{xcolor}
\usepackage{tensor}

\newcommand{\nc}{\newcommand}
\nc{\lb}{\llbracket}
\nc{\rb}{\rrbracket}
\nc{\gl}{\llbracket}
\nc{\gr}{\rrbracket}
\nc{\del}{\partial}
\nc{\tri}{\hspace{-3.5pt}\vartriangle\hspace{-3.5pt}}
\nc{\blacktri}{\blacktriangle}

\nc{\eq}[1]{\begin{equation}
                     \begin{split} #1 \end{split}
                     \end{equation}}
\nc{\ul}{\underline}
\nc{\ov}{\overline}

\nc{\fa}{\hat}
\nc{\fb}{\MakeUppercase}
\nc{\fc}{\tilde }
\nc{\Lie}{{\cal L}} 
\nc{\lambdabar}{{\mkern0.75mu\mathchar '26\mkern -9.75mu\lambda}}

\allowdisplaybreaks[2]
\numberwithin{equation}{section}

\DeclareUnicodeCharacter{2212}{-}
\begin{document}

\vspace*{-1.5cm}
\begin{flushright}
  {\small
  MPP-2021-201\\
  }
\end{flushright}

\vspace{1.5cm}
\begin{center}
  {\Large
    Open-Closed Correspondence of K-theory and Cobordism
} 
\vspace{0.4cm}

\end{center}

\vspace{0.35cm}
\begin{center}
Ralph Blumenhagen and
Niccol\`o Cribiori
\end{center}

\vspace{0.1cm}
\begin{center} 
\emph{
Max-Planck-Institut f\"ur Physik (Werner-Heisenberg-Institut), \\[.1cm] 
   F\"ohringer Ring 6,  80805 M\"unchen, Germany } 
   \\[0.1cm] 
 \vspace{0.3cm} 
\end{center} 

\vspace{0.5cm}

\begin{abstract}
  \noindent
  Non-trivial  K-theory groups and non-trivial cobordism
  groups can lead to global symmetries which are conjectured to be absent
  in quantum gravity.
  Inspired by open-closed string duality, we propose a correspondence
  between the two groups, which can be considered as the physical
  manifestation of a generalisation of the classic Conner--Floyd isomorphism.
  The picture is exemplified by the relations between 
  KO-groups and Spin-cobordisms and between K-groups
  and Spin$^c$-cobordisms.
  Global symmetries related by such isomorphism are
  eventually gauged. By combining K-theory 
  and cobordism, we recover then tadpole cancellation conditions in type I string
  theory and F-theory from a bottom-up perspective.
\end{abstract}

\thispagestyle{empty}
\clearpage

\tableofcontents

\section{Introduction}
\label{sec:intro}

In the swampland program (see \cite{Palti:2019pca,vanBeest:2021lhn} for reviews), the absence of 
global symmetries in quantum gravity (QG) plays a central
role. A recent incarnation of this is the so-called cobordism
conjecture  \cite{McNamara:2019rup},  based on the observation
that a non-vanishing cobordism group would lead to
a global symmetry. Therefore, the conjecture says
that any physically consistent configuration in quantum gravity
must not carry any cobordism charge.\footnote{Note that cobordism groups also
  play a prominent role in the computation of
  Dai-Freed anomalies, see e.g.~\cite{Garcia-Etxebarria:2018ajm,Debray:2021vob}.}

This is quite a generic proposal and, since it is not clear yet what
the correct QG-structure is, it calls for more non-trivial tests and
for relations to other maybe better understood properties of string theory.
First, the conjecture is of topological nature and as such it does not rely
on supersymmetry, the structure on which many
established results in string theory are based.
Second, in view of the fact that string theory, as we know it, is a background
dependent formulation of QG, it is remarkable that due to the very definition
of cobordisms even topology changing configurations
are captured. Therefore, the conjecture really involves processes that one
would expect in quantum gravity, though describing them
(only) at the topological level. Third, the cobordism 
groups of purely geometric structure, like e.g.~$\Omega^{\rm Spin}$, which are studied in the
mathematical literature, are expected to be related to the closed string (gravity) sector
of the theory.  Indeed, strong results have recently been derived by exploiting the cobordism
conjecture (for Pin$^\pm$-structure) in theories with 16 supercharges
and a  small number of compact dimensions \cite{Montero:2020icj,Hamada:2021bbz,Bedroya:2021fbu}. 
This lent support for the so-called string lamppost principle.

A much better studied fundamental principle of quantum gravity is
holo\-graphy, as it is manifest e.g.~in the AdS/CFT correspondence.
In QG, there is a deep connection between gauge  theories on the
boundary and gravity  theories in the bulk, which on the level of the string world-sheet
is known as open-closed string duality. For example, a D-brane can be described
by open strings ending on it, or equivalently as a coherent (boundary) state
of closed string excitations. This is not
just a qualitative picture, but is realised on the quantitative level, e.g.~in the computation of  annulus diagrams.
Thus, one may ask whether there is a reflection of this duality into the cobordism conjecture.

Indeed, there is an apparent very similar open string story, namely
the classification of D-branes via K-theory \cite{Witten:1998cd}. In this context, one talks
about topologically equivalent vector bundles over manifolds
wrapped by D-branes, and  supersymmetry is not important either.
In fact, the starting point is actually given by non-supersymmetric
brane-antibrane pairs equipped with vector bundles.
Moreover, vector bundles of different rank are in the same
K-theory equivalence class, i.e. the open string configuration
is allowed to undergo dramatic topology changes.
An interesting question in this context is whether the K-theory charge
need to vanish on a compact space.\footnote{It was pointed out to us by the referee that \cite{Freed:2000tt}
  implies that in fact all K-theory charges are gauged.} In \cite{Uranga:2000xp}, a brane probe
argument was given that suggested it, while in \cite{Blumenhagen:2019kqm} the
non-cancellation of the K-theory charge was related to a violation of the de Sitter swampland
conjecture  (see also \cite{Damian:2019bkb}).\footnote{In \href{https://indico.cern.ch/event/782251/contributions/3441908/attachments/1871074/3078899/Garcia_Etxebarria.pdf}{his talk} at the ``String Phenomenology  2019''
  conference, I\~{n}aki Garc\'ia-Etxebarria was arguing that in the type I string a non-vanishing K-theory
  charge (for the non-BPS D7-brane) actually leads to a Dai-Freed anomaly. 
   }
If the K-theory charge is a global charge, it should better vanish in QG. Although one might object that K-theory is defined
for branes carrying vector bundles on fixed background spaces,
and thus gravity is actually not manifest.

It is the purpose of this work to argue that
open-closed string duality suggests a correspondence between (open string) K-theory groups
and appropriate (closed string) cobordism groups.
In sections \ref{sec_Ktheory} and \ref{sec_cobord}, we set the stage by briefly reviewing the salient
features of the just mentioned structures and their relevance
in string theory. In section \ref{sec_KCisom}, we first provide an
heuristic  physical
reason why we expect K-theory and cobordism to be related and then 
search the mathematical literature to find
a confirmation in terms of a classic theorem, called
the Conner--Floyd theorem. We will see that for 
D-branes in type I and type II string theory, a generalisation
of this theorem is needed that was formulated and proven
by Hopkins--Hovey in 1992.

Being now related to cobordisms, in section \ref{sec_tad}
we analyse what the cobordism conjecture
implies for D-brane configurations carrying non-vanishing K-theory charge.
A global symmetry in quantum gravity must either be broken or gauged.
We propose that the physical significance of the Hopkins--Hovey
theorem is that it provides the setting where global symmetries 
induced by non-vanishing cobordism/K-theory groups are gauged by continuous or 
discrete RR forms.
We show that tadpoles cancellation conditions
of type I and type IIB orientifold/F-theory compactifications  can be derived
from a bottom-up perspective,
by exploiting the proposed correspondence between cobordism and K-theory.

This makes it evident that K-theory and cobordism 
are like two sides of the same coin, the
first giving the D-brane contribution and the second the topological
bulk contribution to tadpole cancellation conditions. From this
perspective, the initial appearance of global symmetries on each side is
just an artifact of ignoring the gauge field and looking at an
incomplete set of charged objects.

\section{D-branes and K-theory}
\label{sec_Ktheory}

Since the work \cite{Witten:1998cd}, it is a well established fact that 
D-branes in string theory are not classified by homology but by
K-theory. In type II string theory this difference is not really
visible, but for the type I string it becomes apparent.
Indeed, in type I it is known that there exist boundary states in CFT
that break supersymmetry but are still free of any tachyonic
instability, at least as long as one has only one single such non-BPS
brane. Placing two branes on top of each other leads to a tachyonic
mode inducing an instability of the system.
As pointed out by Witten, this behaviour is captured by a non-trivial
$\mathbb Z_2$-valued K-theory class.

Concretely, the stable $Dp$-branes of the $d=10$ type I string are given by the
(reduced) K-theory groups $\widetilde{KO}(S^n)$ with $n=d-p-1$.
These are listed in table \ref{table:koclass}, from which the appearance of stable
non-BPS branes is evident. They are related to $\mathbb Z_2$-valued
reduced K-theory classes. 

\begin{table}[h] 
  \renewcommand{\arraystretch}{1.5} 
\begin{center}
\begin{tabular}{c | c c c c c c c c c c c}
  n & 0 & 1 &  2 & 3 &4 & 5 & 6 & 7 & 8 & 9 & 10 \\
  \hline 
  $\widetilde{KO}(S^n)$ & $\mathbb{Z}$  &  $\mathbb{Z}_2$ &  $\mathbb{Z}_2$
    & 0  &  $\mathbb{Z}$ & 0 & 0 & 0&   $\mathbb{Z}$ &  $\mathbb{Z}_2$&  $\mathbb{Z}_2$\\[0.1cm]
  D-brane & $D9$ & $\widehat{D8}$ & $\widehat{D7}$ & $\widehat{D6}$ &  $D5$
    & $\widehat{D4}$ & $\widehat{D3}$ & $\widehat{D2}$ & $D1$  &$\widehat{D0}$ &$\widehat{D(-1)}$\\
\end{tabular}
\caption{K-theory classes and type I D-branes. Non-BPS D-branes are indicated with a hat.}
\label{table:koclass} 
\end{center}
\end{table}

One can think of these classes as describing bound states of
$D9$-$\ov{D9}$ branes (with non-trivial vector bundles), upon wrapping them on a compact sphere $S^n$.
The reduced K-theory group appears when one considers only initial
$D9$-brane tadpole cancelling configurations,
i.e. $N_{D9}-N_{\ov{D9}}=0$, ignoring
the always present 32 $D9$-branes in the background.
This means that also potential tachyons arising
from open strings stretched between these 32 $D9$-branes and
the condensing  $D9$-$\ov{D9}$ brane pairs are not captured.
The latter indeed appear for the non-BPS $\widehat{D8}$- and
$\widehat{D7}$-branes (see e.g. \cite{Loaiza-Brito:2001yer} for fate of this brane).

For the type IIB string, D-branes are classified by the unitary
reduced K-theory classes $\tilde K(S^n)$. Given that $\tilde
K(S^0)=\mathbb Z$ and $\tilde K(S^1)=0$, and by exploiting Bott
periodicity, one finds the usual D-brane spectrum without any extra
stable non-BPS brane. The story for type IIA D-branes is similar, even
if a bit more subtle.\footnote{To avoid confusion, let us emphasize
  that the relation to Spin$^c$ cobordism that we propose in the
  following section is only meant for the type IIB case. The type IIA case is not yet understood.} It turns out \cite{Witten:1998cd,Horava:1998jy} that they are classified by $\widetilde{K_1}(S^n) = \widetilde{K}(S^{n+1})$ and the full spectrum can be recovered again thanks to Bott periodicity. 
The additional dimension is reminiscent of M-theory, even if a clear connection is not obvious to us\footnote{Recently, in \cite{Sati:2021uhj} charge quantisation in M-theory has been related to cobordism. This is further motivated by several non-trivial consistency checks \cite{Fiorenza:2019usl,Sati:2019tqq,Sati:2020cml}.}.
Type II D-branes and the corresponding K-theory groups are listed in table \ref{table:kclass}.

\begin{table}[ht] 
  \renewcommand{\arraystretch}{1.5} 
\begin{center}
\begin{tabular}{c | c c c c c c c c c c }
  n & 0 & 1 &  2 & 3 &4 & 5 & 6 & 7 & 8 & 9  \\
  \hline 
  $\widetilde{K}(S^n)$ & $\mathbb{Z}$  &  0 &  $\mathbb{Z}$
    & 0  &  $\mathbb{Z}$ & 0 & $\mathbb{Z}$ & 0&   $\mathbb{Z}$ &  0 \\[0.1cm]
  D-brane & $D9$ & - & $D7/D8$ & - &  $D5/D6$
    & - & $D3/D4$ & - & $D1/D2$  & -\\
\end{tabular}
\caption{K-theory classes and type II D-branes. There are no stable non-BPS D-branes.}
\label{table:kclass} 
\end{center}
\end{table}

For later purposes, let us recall here a bit more about K-theory.
One defines the higher reduced K-theory groups of vector bundles over
a manifold $X$ as
\eq{
             \widetilde K_n(X):= \widetilde K(\Sigma_n X),
           }
where $\Sigma X$ is the reduced suspension of $X$, which is
homeomorphic to the smash product of $X$ with $S^1$,
\begin{equation}
\Sigma X \cong S^1 \wedge X , \qquad \Sigma_n X \cong S^n \wedge X.
\end{equation}
Employing the relation $K(X)=\widetilde K(X\sqcup pt)$ and recalling that the $0$-sphere $S^0$ consists of two points, one consequently has
\begin{equation}
\label{KtotildeK}
K_n(pt) =  \widetilde K_n(pt\sqcup pt) = \widetilde K_n(S^0)  =\widetilde K(\Sigma_n S^0)= \widetilde K (S^n)\,,
\end{equation}
and similarly for the $\widetilde{KO}$ groups.
By considering not only direct sums but also tensor products of vector bundles, $K_n(pt)$
is actually a ring.

It is known that the usual BPS $Dp$-brane is coupled to a higher-form RR gauge field $C_{p+1}$, enjoying
a continuous $p$-form  gauge symmetry $C_{p+1}\to  C_{p+1}+d\lambda_p$. As a consequence, 
there will be a conserved $(p+1)$-form (electric) current satisfying
\eq{
  \label{pformeom}
  d\star F_{p+2}=\star J_{p+1}.
  }
  That the current is conserved follows from the fact that it is co-closed, $d \star J_{p+1}=0$.
For a static configuration, the current
can be written as
\eq{
  \star J_{p+1}=\sum_i Q_i\, \delta^{(d-p-1)}(\Delta_{p+1,\, i}),
   \label{currentopen}
}
where the sum is over all $Dp$-branes
carrying charge $Q_i$. 
The $\delta$-function on the right-hand-side
is the Poincare dual of the $(p+1)$-cycle $\Delta_{p+1,\, i}$ wrapped by
the brane.
Integrating \eqref{pformeom} over a compact space implies that the sum
over all charges on it has to vanish.

Non-BPS $\widehat{Dp}$-branes are not coupled to higher $(p+1)$-form RR gauge fields, but one
can still assign a $\mathbb Z_k$ K-theory charge to them. 
The framework of differential K-theory \cite{Freed:2000tt,Freed:2000ta} suggests 
to associate to this charge a 
 discrete gauge symmetry, 
such that the total charge would vanish when integrated over a compact
manifold.\footnote{We thank the referee for pointing this out to us. We
also thank I\~{n}aki Garc\'ia-Etxebarria for discussions on this point.}

It is one of the most fundamental swampland conjectures
that in quantum gravity there should be no global symmetries,
i.e.~they should either be broken or gauged. Therefore,
one can consider a non-vanishing K-theory class as a topological  obstruction
of having such an object in a string theory background.
For instance, placing a single non-BPS $\widehat{D7}$-brane on $S^2$
has  a $\widetilde{KO} (S^2)=\mathbb Z_2$ obstruction and therefore
would not be allowed.

These arguments sound very similar to the recent proposal
of a vanishing cobordism class in quantum gravity \cite{McNamara:2019rup}. 
In fact the purpose of this work is to present a physical reason for a deep connection
between the two. Before we discuss such a connection, let us review
the cobordism conjecture.

\section{Global Symmetries and Cobordism Groups}
\label{sec_cobord}

The set-up is very similar to that of the last section, however
there are a priori no branes involved. One considers $d$-dimensional
string theory (i.e.~$d=10$) and places it on an $n$-dimensional
compact space $M$, thus leaving $D=d-n$ non-compact dimensions.
In \cite{McNamara:2019rup}, it was argued that a non-trivial cobordism
group $\Omega_n^{\widetilde{QG}}$ of quantum gravity configurations with $n$ compact dimensions
implies a $(d-n-1)$-form global symmetry. Since there should be
no global symmetries, a non-vanishing cobordism class is
an obstruction to a valid quantum gravity background.
Thus, if possible, such a  background should be extended to remove the
obstruction and arrive  at vanishing cobordism classes.
This can be achieved in two ways:
\begin{itemize}
 \item {\bf Breaking of the symmetry}: it is obtained by introducing 
 non-trivial $(d-n-1)$-dimensional defects so that the global current is not conserved, i.e.  
\eq{
 0\neq d \star J_{d-n}= I_{n+1}=\sum_{{\rm def},j} \delta^{(n+1)}(\Delta_{d-n-1,\, j})\,,
}
where the $\delta$-functions are the Poincare dual of the
$(d-n-1)$-cycles wrapped by the defects.
Then, elements in the kernel of the map
 \eq{
      \Omega_n^{\widetilde{QG}}\to \Omega_n^{\widetilde{QG}+{\rm defects}}
 }
 are killed in the full theory. Here, the refined cobordism group on
 the right hand side is obtained from the original cobordism
 by mapping onto those manifolds which do not include defects. 

\item {\bf Gauging of the symmetry}: it means that the cobordism class
    of a consistent configuration is the 0-element, i.e. $[M]=0\in
    \Omega_n^{\widetilde{QG}}$. In other words, the total charge on a
    compact manifold has to vanish. One says that  elements in the cokernel of the map
\eq{
     \Omega_n^{\widetilde{QG};\,{\rm gauging}}\to \Omega_n^{\widetilde{QG}}
 }
 are co-killed. Here,
the cobordism for the refined approximation maps into the original
cobordism by  forgetting  the  gauge  field.   
 
\end{itemize}

The first option is discussed at length in \cite{McNamara:2019rup} and
has been made further explicit in concrete non-supersymmetric
string backgrounds with  uncanceled NS-NS tadpoles in
\cite{Buratti:2021yia,Buratti:2021fiv}.  In a few words, taking the backreaction of such
configurations into account, one gets a spontaneous compactification
of initially flat directions \cite{Dudas:2000ff,Blumenhagen:2000dc}  of   finite size, whose boundary
then supports the required defect. Besides, one could wonder if and
how the cobordism group, which is abelian by construction, can detect
non-abelian defects. For the prototype example of $(p,q)$ 7-branes in
F-theory, this puzzle has been resolved in \cite{Dierigl:2020lai} 
  (see also \cite{McNamara:2021cuo} for a more general discussion).

The focus of the present work is on the second option, namely the gauging
of the symmetry. Indeed, we will show that by exploiting the proposed
correspondence between cobordism and K-theory we
are led
to known tadpole constraints in string theory.

\subsection{Spin Cobordism}
  
The main question of course is what the full $\Omega_n^{QG}$ actually is.
In mathematics, one usually considers only cobordism groups of
geometric structure, where
the manifolds in question are often taken from a special class.
Some of them have been discussed in \cite{McNamara:2019rup} as well.
For instance, one can consider cobordisms for Spin-manifolds, i.e.~oriented 
manifolds on which one can globally define (uncharged) fermions.
In this case, one says that two $n$-dimensional Spin-manifolds $M$ and $N$ are
cobordant if there exists an $(n+1)$-dimensional Spin-manifold $W$ such that
$\partial W=M\sqcup \ov N$, where $\ov N$ is the orientation reversed manifold.
The  equivalence classes $[M]$ make up the Spin-cobordism group $\Omega_n^{\rm
  Spin}$, where the addition is defined via disjoint unions,
i.e.~$[M]+[N]=[M\sqcup N]$. Similarly to K-theory, via the cartesian product
of two manifolds, $M\times N$, the Spin-cobordism theory
\eq{
  \Omega_*^{\rm Spin}=   \bigoplus_{n=0}^\infty \Omega_n^{\rm Spin}
}
can be given the structure of a ring.

The cobordism groups $\Omega_n^{\rm Spin}$ and their
generators $\Sigma_{n,i}$ for $0\le n\le 8$ are listed in table \ref{table:bordclassspin}.

\vspace{-0.0cm}
\begin{table}[ht] 
  \renewcommand{\arraystretch}{1.5} 
\begin{center}
\begin{tabular}{c | c c c c c c c c c c c}
  n & 0 & 1 &  2 & 3 &4 & 5 & 6 & 7 & 8 \\
  \hline 
  $\Omega_n^{\rm Spin}$& $\mathbb{Z}$  &  $\mathbb{Z}_2$ &  $\mathbb{Z}_2$
    & 0  &  $\mathbb{Z}$ & 0 & 0 & 0&   $\mathbb{Z}^2$ \\[0.1cm]
  $\Sigma_{n,i}$ & pt$^+$ & $S^1_{p}$ & $S^1_{p}\times S^1_{p}$ & 0 & $K3$ & 0 & 0 & 0 & $\mathbb{B}\oplus \mathbb{HP}^2$ \\
\end{tabular}
\caption{Cobordism classes for Spin-manifolds and their generators $\Sigma_{n,i}$. Here, by
  $\mathbb{Z}^n$ we mean $\mathbb{Z}\oplus \ldots \oplus\mathbb{Z}$,
  $n$ times. pt$^{+}$ is the positive oriented point, $S^1_{p}$ indicates the circle with periodic boundary
  conditions for the fermions, $\mathbb{B}$ the so-called Bott space and $\mathbb{HP}^2$ is the hyperbolic projective space.}
\label{table:bordclassspin} 
\end{center}
\end{table}

\vspace{-0.7cm}

\subsection{Spin$^c$  Cobordism}

Analogously, one can introduce the
cobordism ring for Spin$^c$-manifolds, i.e.~oriented manifolds on
which charged fermions can be globally defined (see e.g. \cite{Freed:1999vc}). 
To make chiral indices well-defined, one needs an
uplift of the second Stiefel-Whitney class from $\mathbb Z_2$-cohomology 
to  $\mathbb Z$-valued cohomology. In other words, a
complex Spin$^c$-manifold $\Sigma$ also carries a line-bundle ${\cal L}$ so
that
\eq{
                  c_1({\cal L})-{1\over 2} c_1(\Sigma) \in    H^2(\Sigma,\mathbb Z)\,.
}                   
For $\Sigma$ not Spin, a half-integer gauge flux can still make it Spin$^c$.
Clearly, every Spin-manifold is also a Spin$^c$-manifold upon choosing
the trivial line bundle, ${\cal L}={\cal O}$.

The cobordism groups for ${\rm Spin}^c$ are listed
in \cite{McNamara:2019rup}, where up to $n=4$ also the generators are given.
In order to extend that list to $n=6$, let us recall how one finds
the two generators of $\Omega_4^{{\rm Spin}^c}=\mathbb Z\oplus \mathbb
Z$. These can be found  by first noticing that the two cobordism invariants in this case are
the Todd class ${\rm td}_4=(c_2(\Sigma) +c_1^2(\Sigma))/12$ and $c_1^2(\Sigma)$ \cite{Wan:2018bns} .
The question then is how to identify two complex
twofolds generating the full charge lattice $\mathbb Z\oplus
\mathbb Z$. This is not too difficult, as by scanning some simple projective spaces
one finds

\vspace{0.7cm}
\begin{table}[h] 
  \renewcommand{\arraystretch}{1.5} 
  \begin{center}
 \vspace{-0.7cm}
\begin{tabular}{c |   c c }
  space &  td$_4$  & $c_1^2$   \\
  \hline
  ${\mathcal M}_1=\mathbb{P}^2$ & 1 & 9\\
  ${\mathcal M}_2=\mathbb{P}^1\times \mathbb{P}^1 $ &  1 & 8\\   
\end{tabular}
\vspace{-0.6cm}
\end{center}
\end{table}

\noindent
Therefore,  $[\mathcal{G}_1]=-8[{\mathcal M}_1]+9[{\mathcal M}_2]$
carries charge $(1,0)$ and 
$[\mathcal{G}_2]=[{\mathcal M}_1]-[{\mathcal M}_2]$  charge $(0,1)$
so that they can be identified as  the diagonal generators
of $\mathbb Z\oplus \mathbb Z$. Note that in cobordism one has
$[\mathcal{G}_1]=[dP_9]$, with $dP_9$ being the rational elliptic surface.
For $n=6$, we show in appendix \ref{app_todd6} that  the two cobordism invariants listed in
\cite{Wan:2018bns} are
\eq{
  {\rm td}_6=c_2(\Sigma) \,c_1(\Sigma) /24\,,\qquad  c_1^3(\Sigma) /2 \,.
} 
We notice that the projective spaces $\{\mathbb{P}^2\times \mathbb{P}^1,(\mathbb{P}^1)^3\}$
only generate a $\mathbb Z\oplus3\mathbb Z$ sublattice. 
We found instead that the
following simple non-toric spaces work.

\begin{table}[h!] 
  \renewcommand{\arraystretch}{1.5} 
  \begin{center}
    \vspace{-0.3cm}
\begin{tabular}{c |   c c }
  space &  td$_6$  & $c_1^3/2$   \\
  \hline
  ${\mathcal N}_1=\begin{matrix} \mathbb{P}^2
                 \\[-0.4cm] \mathbb{P}^2 \end{matrix}
               \bigg[ \begin{matrix} 2 \\[-0.4cm] 2 \end{matrix}\bigg]$ & 1 & 6\\
  ${\mathcal N}_2=\begin{matrix} \mathbb{P}^3
                 \\[-0.4cm] \mathbb{P}^1 \end{matrix}
               \bigg[ \begin{matrix} 3 \\[-0.4cm] 1 \end{matrix}\bigg]$ &  1 & 5\\   
\end{tabular}
\vspace{-0.7cm}
\end{center}
\end{table}

\noindent
Therefore,  $[\mathcal{H}_1]=-5[{\mathcal N}_1]+6[{\mathcal N}_2]$
carries charge $(1,0)$ and 
$[\mathcal{H}_2]=[{\mathcal N}_1]-[{\mathcal N}_2]$  charge $(0,1)$,
so that they can be identified as  the diagonal generators
of $\mathbb Z\oplus \mathbb Z$. A connected representative for
 $[\mathcal{H}_1]$ is the threefold $[dP_9\times \mathbb P^1]$.
Up to $n=6$, we  arrive at the cobordism classes 
shown in table \ref{table:bordclassspinc}.

\begin{table}[h] 
  \renewcommand{\arraystretch}{1.5} 
  \begin{center}
     \vspace{-0.2cm}
\begin{tabular}{c |     c c c c c c }
  n & 0  &  2  &4  & 6   \\
  \hline 
    $\Omega_n^{{\rm Spin}^c}$& $\mathbb{Z}$  &   $\mathbb{Z}$ 
    &  $\mathbb{Z}^2$ & $\mathbb{Z}^2$  \\
   $\Sigma_{n,i}$ & pt$^+$ &  $\mathbb{P}^1$ &  $\mathbb{P}^2\oplus
                                               (\mathbb{P}^1)^{2} $ &
                                                                     $\begin{matrix} \mathbb{P}^2
                 \\[-0.4cm] \mathbb{P}^2 \end{matrix}
               \bigg[ \begin{matrix} 2 \\[-0.4cm]
                 2 \end{matrix}\bigg]\oplus \begin{matrix} \mathbb{P}^3
                 \\[-0.4cm] \mathbb{P}^1 \end{matrix}
               \bigg[ \begin{matrix} 3 \\[-0.4cm]
                 1 \end{matrix}\bigg]$ \\
    \end{tabular}
\caption{Cobordism classes and generators $\Sigma_{n,i}$ for Spin$^c$ manifolds. 
  Here,  $(\mathbb{P}^k)^{l}=\mathbb{P}^k\times\ldots\times
    \mathbb{P}^k$,  $l$ times and the line bundle is generically chosen
    as ${\cal L}^2=H_\Sigma$, with $H_\Sigma$ denoting the hyperplane bundle.}
    \label{table:bordclassspinc}
     \vspace{-0.9cm}
\end{center}
\end{table}

Comparing them to the K-theory classes from the previous section,
there is a striking similarity between $\widetilde K(S^n)$ and
$\Omega_n^{{\rm Spin}^c}$ as well as between $\widetilde{KO}(S^n)$
and $\Omega_n^{{\rm Spin}}$: the same classes are vanishing and also
the same $\mathbb Z_k$ factors appear. It is only the number of
$\mathbb Z_k$-factors that differs for a couple of cases. In particular, from the tables one deduces that $\widetilde{KO}(S^n) \subseteq \Omega_n^{{\rm Spin}} $ and $\widetilde K(S^n) \subseteq \Omega_n^{{\rm Spin}^c}$. 
If there is such a relation between K-theory and cobordism, then one
might expect it to be known in the mathematical literature. Indeed,
in section \ref{sec_KCisom} we will first present a physical argument for such
a relation and then state a generalisation of the classic Conner--Floyd
theorem, which was formulated and proven by Hopkins--Hovey and 
which provides the expected mathematical relation.

\subsection{String  Cobordism}
\label{subsec_string}

For later purposes, let us also mention the so-called string
cobordism groups, $\Omega_n^{\rm String}$. They are relevant in the
 context   of the heterotic string and for its S-dual type I string.
  One is also including in the structure the Kalb-Ramond $B$-field and its
 non-trivial Bianchi identity
 \eq{
   \label{stringbianchi}
   dH={p_1(\Sigma)\over 2}= -{1\over 16\pi^2} {\rm tr}(R\wedge R),
 }  
 where $p_1(\Sigma)$ denotes the first Pontryagin
 class  of the Spin manifold $\Sigma$. A potential gauge field
 contributing like $p_1(F)$ is set to zero in \eqref{stringbianchi}.
 As a consequence of the Bianchi identity, the theory can only be well-defined  
 for a trivialisation of $p_1/2$, i.e.~one only includes
 manifolds satisfying   $p_1(\Sigma)/2=0$ in cohomology. This is referred
 to as a string structure and the corresponding  cobordism groups, $\Omega_n^{\rm String}(pt)$,
 are called string cobordism.  In table \ref{table:bordclassstring}, we list the lower dimensional
 string cobordism groups together with their generators \cite{McNamara:2019rup}.
  
\vspace{-0.2cm}
\begin{table}[ht] 
  \renewcommand{\arraystretch}{1.5} 
\begin{center}
\begin{tabular}{c | c c c c c c c c c}
  n & 0 & 1 &  2 & 3 &4 & 5 & 6  \\
  \hline 
  $\Omega_n^{\rm String}$& $\mathbb{Z}$  &  $\mathbb{Z}_2$ &  $\mathbb{Z}_2$
    & $\mathbb{Z}_{24}$  &  0 & 0 &  $\mathbb{Z}_2$  \\[0.1cm]
  $\Sigma_{n,i}$ & pt$^+$ & $S^1_{p}$ & $S^1_{p}\times S^1_{p}$ & $S^3_{H}$ & 0 & 0 &  $S^3_{H}\times S^3_{H}$  \\
\end{tabular}
\caption{Cobordism classes for String-manifolds and their generators
  $\Sigma_{n,i}$.  Here, $S^3_{H}$ means that the $S^3$ support a unit three-form
flux $H$.}
\label{table:bordclassstring} 
\end{center}
\end{table}

\vspace{-0.6cm}
\noindent
The interplay between cobordism groups and non-trivial Bianchi identities 
will be developed further in section \ref{sec_tad}.

\section{K-theory -- Cobordism  Isomorphism}
\label{sec_KCisom}

First, let us present  an  intuitive physical 
argument for the existence of the aforementioned relation.
We stress that this is an heuristic picture, 
and as such should not be overestimated, nevertheless it will put us
on the right track.

Recall that D-branes can be seen in two ways, as it
is also explicit in their CFT description. One can either consider
them as the end-points of open strings, whose quantisation
describes fluctuations of the D-branes giving rise
to massless gauge fields. This picture is clearly the one 
underlying the K-theory classification of D-branes, which is about
equivalence classes of vector bundles on manifolds wrapped by the
branes.

However, D-branes can also be considered as boundary states
in the closed string Hilbert space, which makes it manifest that
they couple to closed string excitations. At the massless level, this
means that D-branes backreact on the geometry, the dilaton  and also
the other $p$-form fields present in effective string theories.
Thus, D-branes can be considered as specific sourced  solutions in
  supergravity. However, it is well known that a $Dp$-brane can also
  be considered as a solitonic  magnetically
  charged object.  As such, it carries a topological obstruction against decaying to
the vacuum. In the following, we will refer to this as the (dual) bulk picture of a $Dp$-brane.

Due to open string - closed string duality, also called
loop channel - tree channel equivalence in the CFT context, both
descriptions should be equivalent. In our framework, this implies
  that  for every obstruction to the decay of a (non-BPS) $Dp$-brane to the vacuum, classified e.g.~in type I by  $\widetilde{KO}(S^n)$ with
  $n=d-p-1$,  there should exist  an equivalent  obstruction in  the dual
  bulk picture of the brane. The observed equivalence of
  K-theory and cobordism suggests that this obstruction is described 
  by an  appropriate non-vanishing cobordism class
$\Omega^{\widetilde{QG}}_n$. Consistently, both obstructions give rise to a global
$(d-n-1)$-form symmetry.
As alluded to previously, this picture is of course a simplification
and we will see later that one actually needs more than a
single D-brane or O-plane to end up with a compact bulk manifold representing a given cobordism class.

Since for the construction of the $\widetilde{KO}(S^n)$  classes
the 32 background  $D9$-branes were ignored, 
for a (non)-BPS brane in type I we expect the
backreaction to only involve the closed string fields.
The fermions, i.e.~the gravitinos and dilatinos, are neutral 
so that, upon ignoring the RR fields and the dilaton, it is
suggestive  to consider the Spin-cobordism
groups $\Omega^{\rm Spin}_n$, at least in first approximation.\footnote{By
  ignoring  the RR two-form field $C_2$,  at this stage we are not
  sensitive to  potential  issues with the type I Bianchi identity
  $dF_3\sim {\rm tr}(R\wedge R)-{\rm tr}(F\wedge F)$. The latter would
  suggest to look at $\Omega^{\rm String}_n$ rather than $\Omega^{\rm
    Spin}_n$.}
On the other hand, it was pointed out in \cite{McNamara:2019rup} that F-theory
compactifications on elliptically fibered Calabi-Yau spaces induce a
Spin$^c\!$-structure on the base manifolds.\footnote{The actual
  relation of F-theory and Spin$^c$ is more involved, as the relevant
  groups are
  cobordism groups of a manifold with an elliptic fibration, which are
  difficult to compute.}
This suggests a relation between the branes classified by
$\widetilde{K}(S^n)$ classes and Spin$^c\!$-cobordism groups, $\Omega^{{\rm
    Spin}^c}_n$.

This intuitive reasoning supports the existence of a map relating
cobordism to K-theory, with the following consequences:
\begin{itemize}
\item{If there exists a $\mathbb Z_k$ obstruction in some K-theory
class  $\widetilde{KO}(S^n)$, then also $\Omega^{\rm Spin}_n$ should
contain this obstruction.}
\item{However, $\Omega^{\rm Spin}_n$ could admit  other equivalence
    classes that are not related to D-branes.
      Therefore, one cannot simply conclude that
    $\widetilde{KO}(S^n)$ is isomorphic to $\Omega^{\rm Spin}_n$.
  }
\item{Nevertheless, as for AdS/CFT, one might expect that open-closed
   string duality is a one-to-one map, so that there should exist a way to make
    the above map an isomorphism.}
 \end{itemize}

 \noindent
 In fact, in mathematics a quite advanced framework is known
 that substantiates the above physical intuition.

 \subsection{Atiyah--Bott--Shapiro orientation}

The work of  Atiyah--Bott--Shapiro (ABS)  \cite{Atiyah:1964zz} set the
ground for a relation between Spin-cobordism and K-theory.
In particular, it contains the definition of ring homomorphisms
\eq{
\alpha^c:\Omega^{{\rm Spin}^c}_* \to K_*(pt)\,,\qquad
\alpha:\Omega^{\rm Spin}_* \to KO_*(pt)\,.
}
Let us discuss these two maps in more detail.

\subsubsection*{Definition of $\alpha^c$}

When restricted to a fixed grade $n$, the homomorphism $\alpha^c$
is the Todd genus, i.e.~for an $n$-dimensional manifold $M$ we have
\eq{
  \label{abshomospinc}
  \alpha^c_n([M])= {\rm Td}(M)\in \mathbb Z\,.
}
Later, we will also use that the Todd genus is the integral over the
top Todd class, i.e. ${\rm Td}(M)=\int_M {\rm td}_n(M)$ with ${\rm
  td}_n(M)\in H^n(M)$.
Since it actually acts on equivalence classes, for it to be well defined the Todd
genus should better be a cobordism invariant. Moreover, it satisfies 
${\rm Td}(M\sqcup N)={\rm Td}(M)+{\rm Td}(N)$ and
${\rm Td}(M\times N)={\rm Td}(M)\cdot{\rm Td}(N)$.
Given that the Todd genus evaluates to one on any complex projective space, for the generators of $\Omega^{{\rm Spin}^c}_n$ listed in table \ref{table:bordclassspinc}
one finds $\alpha^c_n([\Sigma_{n,i}])=1$, so that the homomorphism
is surjective. The kernel of the map $\alpha^c_4$ is spanned by the class
$[\mathbb{P}^2]-[\mathbb{P}^1\times \mathbb{P}^1]\in \Omega^{{\rm
    Spin}^c}_4$,
while for the map $\alpha^c_6$ we have seen that it is spanned by 
\eq{
                         [\mathcal{G}_2]=[{\mathcal M}_1]-[{\mathcal M}_2]\in \Omega^{{\rm
    Spin}^c}_6\,.
}

\subsubsection*{Definition of $\alpha$}

The story for $\alpha$ is similar, even if slightly more involved.
Let us first look at $KO_*(pt)$.
As a consequence of Bott periodicity, it  can be formulated as the
ring $\mathbb{Z}[\mathrm{s},\mathrm{k},\mathrm{b}]$ modulo appropriate equivalence relations, namely  \cite{KreckStolz}
\begin{equation}
KO_* (pt) = \mathbb{Z}[\mathrm{s},\mathrm{k},\mathrm{b}] / (2\mathrm{s}, \mathrm{s}^3, \mathrm{k}\mathrm{b}, \mathrm{k}^2 - 4\mathrm{b}),
\end{equation}
where $\mathrm{s}$, $\mathrm{k}$, $\mathrm{b}$ are elements of order $1$, $4$ and $8$ respectively.
An explicit definition of the map $\alpha$ is given in \cite{HITCHIN19741}
\begin{equation}
\alpha_n([M]) = \left\{\begin{array}{ccl}
 \hat A(M) & &n=8m,\\
 \hat A(M)/2   & & n = 8m+4,\\
\text{dim} \, H \quad \,\,\,{\rm mod}\ 2 & & n = 8m+1,\\
\text{dim} \, H^+ \quad {\rm mod}\ 2 & & n = 8m+2,\\
0 & &\text{otherwise},
\end{array}\right.
\end{equation}
where $\hat A(M)$ is the
$\hat A$-genus and $H$ ($H^+$) the space of (positive) harmonic
spinors.

Recall from table \ref{table:bordclassspin} that the generator of $\Omega_1^{\rm
  Spin}=\mathbb{Z}_2$ is
the circle $S^1_{p}$ with periodic boundary condition for fermions.
The generator of $\Omega_2^{\rm Spin} = \mathbb{Z}_2$ is 
$S^1_{p}\times S^1_{p}$.
The generator of $\Omega_4^{\rm Spin}=\mathbb{Z}$ is the Kummer
surface $K3$.
The generators of $\Omega_8^{\rm Spin} = \mathbb{Z} \oplus \mathbb{Z}$
are the hyperbolic projective space $\mathbb{HP}^2$ and the so-called
Bott space $\mathbb{B}$.
Therefore, the action of $\alpha$ on the Spin-cobordism generators is
\eq{
\alpha_n(\Sigma_{n,i})=1\ \ {\rm for}\  n\leq 7\,,\qquad  \alpha_8([\mathbb{B}])=1\,,\qquad
\alpha_8([\mathbb{HP}^2])=0\,.
}
One realizes that $\alpha_n$ is surjective and furthermore the Kernel
is non-trivial only for $n\geq 8$.

As mentioned, open-closed string duality is expected to be a
  one-to-one map so that there should exists a way to promote
  the above homomorphisms to isomorphisms. 
 In case the map $\alpha$ is surjective, one can simply divide by its
 kernel  to get  an isomorphism
\begin{equation}
\Omega_n^{\rm Spin} /{\rm ker}(\alpha) \cong  \widetilde{KO}(S^n)
\end{equation}
and similarly for $\alpha^c$.\footnote{The existence of this
   isomorphism follows trivially from Theorem C and Proposition 3.3 of
   \cite{KreckStolz}  for $X=pt$.}
Hence, we propose that this isomorphism is the mathematically precise
reflection of stringy open-closed duality at the topological level.

\subsection{Hopkins--Hovey isomorphism}

So far, we were considering only D-branes related to the K-theory groups
$K_n(pt)$ and $KO_n(pt)$. However, from our picture of open-closed duality,
one would expect that the relation between K-theory and
cobordism holds for type I or type II D-branes on any compact space
$X$.

Indeed, there exists an extension of the isomorphism from the previous
section
to a more general case, which is known as the Hopkins--Hovey isomorphism.
The latter is a generalisation of a classic theorem by Conner--Floyd
\cite{ConnerFloyd}, which in turn is a special case of the Landweber
exact functor theorem \cite{Landweber1976HomologicalPO}.
To be concrete, Hopkins--Hovey proved that the maps
\begin{align}
\Omega^{\rm Spin}_*(X) \otimes_{\Omega^{\rm Spin}_*} \, KO_* &\to KO_*(X),\\[0.1cm]
\Omega^{\rm{Spin}^c}_*(X) \otimes_{\Omega^{\rm{Spin}^c}_*} \, K_* &\to K_*(X)
\end{align}
are isomorphisms for any topological space $X$
\cite{Hopkins1992}. One says that KO (K) theory is isomorphic to the
extension of scalars of Spin (Spin$^c$) cobordism theory through the
ABS-orientation. The definition of the extension
of scalars is given in appendix \ref{app_eos}.

\begin{figure}[h!]
\begin{center}
\begin{tikzpicture}
\draw [line width=0.3mm] (0,0) arc (0:360:0.3 and 1.3);
\draw [line width=0.3mm,dashed] (7.3,0) arc (0:360:0.3 and 1.3); 
\draw [line width=0.3mm] (7.,1.3) arc (90:-90:0.3 and 1.3); 
\draw [line width=0.3mm](1.8,-3) arc (360:0:-1.5 and .3); 
\draw[rounded corners=50pt,line width=0.3mm](-.3,1.3)--(3.6,1)--(7,1.3);
\draw[rounded corners=50pt, line width=0.3mm](-.3,-1.3)--(3.6,-1)--(7,-1.3);
\draw   [line width=0.3mm](.85,-.0) arc (-80:0:1cm and 0.5cm);
\draw  [line width=0.3mm] (1,0.04) arc (194:92:.7cm and 0.3cm);
\draw  [line width=0.3mm](4.,0.12) arc (175:315:1cm and 0.5cm);
\draw  [line width=0.3mm](5.4,-0.38) arc (-31:180:.7cm and 0.4cm);
\node (a) at (-0.3,0) {$M$};
\node (b) at (7.,0) {$N$};
\node (c) at (3.4,0.4) {$W$};
\node (d) at (3.3,-3) {$X$};
\node (e) at (3.6,-2) {$h$};
\node (f) at (1.3,-2) {$f$};
\node (g) at (5.5,-2) {$g$};
\draw [->]  (-0.2,-1.5) -- (1.7,-2.8) ;
\draw [->]  (3.3,-1.2) -- (3.3,-2.5);
\draw [->]  (6.9,-1.5) -- (4.9,-2.8) ;
\end{tikzpicture}
\caption{Cobordism $(W,h)$ between pairs $(M,f)$ and $(N,g)$.}
\label{refcob}
\end{center}
\end{figure}
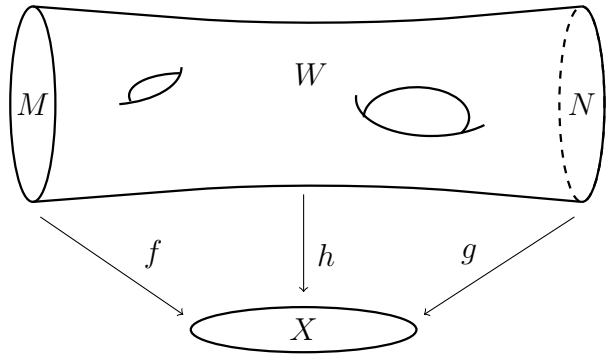

Here, the cobordism groups $\Omega_n^{G}(X)$
are given by equivalence classes of pairs $(M,f)$ of $n$-dimensional manifolds
with $G$-structure together with maps  $f:M\to X$. Two pairs  are equivalent, $(M,f)\sim
(N,g)$, if $M$ and $N$ are cobordant, i.e.~if there exists
an $(n+1)$-dimensional manifold $W$ with $G$ structure such that  $\partial W=M\sqcup \ov N$,
and if there is a map $h:W\to X$, such that $h|_{M}=f$ and $h|_{N}=g$.
 An illustration is given in Figure \ref{refcob}.

Choosing $X=pt$ in the general statement of the Hopkins--Hovey theorem gives back the result from the previous section, now written as the isomorphism
\begin{align}
\Omega_n^{\rm Spin} \otimes_{\Omega_n^{\rm Spin}} \widetilde{KO}(S^n) &\cong \widetilde{KO}(S^n),\\[0.1cm]
\Omega_n^{{\rm Spin}^c} \otimes_{\Omega_n^{{\rm Spin}^c}} \widetilde{K}(S^n)&\cong \widetilde{K}(S^n),
\end{align}
where we used that $KO_n(pt) = \widetilde{KO}(S^n)$, analogously to
\eqref{KtotildeK}.
In this simple case, the isomorphism follows trivially from the  
standard result 
\begin{equation}
\label{RMisom}
R \otimes_R M \cong M,
\end{equation}
which is valid for any ring $R$ (being a module over itself) and any
$R$-module $M$. The short proof is relegated to appendix \ref{app_proof}.

Without pursuing this further at the present stage, we simply conclude  that as duals of
$K_n(X) (KO_n(X))$  the cobordism
groups $\Omega_n^{{\rm Spin}^c}(X) (\Omega_n^{\rm Spin}(X))$ should be relevant in string
theory, as well.

\section{Cobordism, K-theory and Tadpoles}
\label{sec_tad}

As we recalled in section \ref{sec_cobord},
the cobordism conjecture states that the cobordism classes of quantum gravity
must be trivial. This implies that any global symmetry detected by
a non-trivial cobordism class must either be broken by
defects or gauged. 
Since we have now the classical Hopkins--Hovey  theorem available that relates
  certain K-theory classes to certain cobordism groups, we can wonder
  what the physical significance of this relation is.
  What does it mean that e.g. Spin cobordism are related to
  $KO$-groups,  which are relevant for the classification of D-branes in
  type I string theory?  One can certainly also consider Spin
  cobordism in type II string theories, as done in \cite{McNamara:2019rup}. Hence,
  what is special about type I?

   To get an idea, we first observe that K-theory classifies certain
   defects, namely D-branes. As suggested, all global symmetries
   related to K-theory are gauged. Therefore, we propose that the
   Hopkins--Hovey isomorphism provides the relations between the
   D-brane defects and the cobordism groups (charges) that are coupling to the same
   gauge field, i.e.~that appear in the same tadpole cancellation
   condition.

   For instance, as  will be discussed in more detail next, the
   relation between $KO_{4}(pt)=\mathbb Z$ and $\Omega^{{\rm
       Spin}}_{4}(pt)=\mathbb Z$ indicates that the type I D5-branes and the cobordism
   invariant $\hat\alpha_4=\hat A/2$ couple to the RR 6-form gauge field
  and thus both appear in the corresponding tadpole cancellation
  condition. One could also consider $\Omega^{{\rm
       Spin}}_{4}(pt)$ in type IIB theory, but then it is likely not gauged
   but broken \cite{McNamara:2019rup}, as the Hopkins--Hovey isomorphism does not apply to this setup.  In the following we substantiate our proposal by a
   couple of examples.

\subsection{Spin-structure in type I}
\label{subsec_spintad}

The non-vanishing classes $KO_{n}(pt)=\mathbb Z$ ($n=0,4,8$) clearly correspond
to the usual $Dp$-branes of type I with $p=d-n-1$ and $d=10$.
The associated global $p$-form symmetry is gauged,
as there are appropriate RR $(p+1)$-form gauge fields $C_{p+1}$.
Thus,  also  the open-closed dual cobordism group $\Omega^{{\rm
    Spin}}_{n}(pt)$ is expected to be gauged, so that on a compact space only
configurations with vanishing charge are allowed. Let us discuss
this in more detail for a first example with $n=4$, which is central
to the overall picture we intend to convey.

\subsubsection*{Gauging the global symmetry of $\Omega^{{\rm Spin}}_{4}(pt)$}

 For the generator $[K3]\in\Omega_4^{\rm Spin}$,  this is the type I analogue
of the example involving the heterotic string on K3 discussed in
\cite{McNamara:2019rup}.
To understand the origin of the obstruction, we have to bear in mind
that in computing   $\Omega^{{\rm Spin}}_{n}(pt)$ the higher form gauge field
$C_{d-n}$ was set to zero. Then, one finds
a $\mathbb Z$-valued obstruction in this truncated type I theory,
meaning  that the generator $[K3]$ and also all other non-trivial
representatives $[M]$ are  not  topologically consistent solutions.

As we have seen, the ABS-orientation involves the map $\alpha_n$ between cobordism
and K-theory. Therefore, the natural candidate for the global
  current is
  \eq{
    \label{currentbord1}
    \star J_{6}\sim\hat\alpha_4(K3),
  }
where the four-form $\hat\alpha_4\in H^4(K3)$ is such that $\int_{K3}
\hat\alpha_4(K3)=\hat A(K3)/2=\alpha_4([K3])$.
Thus, the current is preserved,  $d\star J_{6}=0$, and 
the charge of $K3$ is given by
  \eq{   
    \int_{K3} \star J_{6}\sim\int_{K3} \hat\alpha_4(K3) =\alpha_4([K3])=1\,.
  }
From \eqref{currentbord1}, it is suggestive that $\hat\alpha_4$ should
be considered as a magnetic (non-singular) current.
Hence, after gauging one would obtain a magnetic Bianchi identity of the type (up to normalisation)
  \eq{
  \label{Bianchiid}
    d\tilde F_{3}=\tilde J_{4}\sim\hat\alpha_4(K3)\,.
    }
     This encodes cobordism information only and thus it is not yet the end of the story, as we are now going to argue.
 
 So far, we have considered sources of the $C_2$-form and of the dual
 $C_6$-form separately on the
 K-theory side (i.e.~$D5$-branes) and on the cobordism side (i.e.~$K3$).
 We propose that the existence of the isomorphism between K-theory and cobordism
 suggests that upon gauging the two initially appearing global symmetries actually couple
 to one and the same $C_2$-form (or its electric dual $C_6$-form). Hence,
 combining them with an appropriate normalisation,
 we have a total Bianchi identity
 \eq{
  \label{Bianchiidtot}
    d\tilde F_{3}\sim \sum_i Q_{i}\, \delta^{(4)}(\Delta_{6,i})-k\,\hat\alpha_4(M)\,
  }
  with $Q_i=\pm 1$ and where we considered general elements $[M]\in
  \Omega_4^{\rm Spin}$.
  To warrant e.g.~anomaly cancellation we know that
  in string theory one has $k=24$.
  In fact, there is an interesting story behind the relative normalisation $k=24$, on which we will comment in a while.
Integration of \eqref{Bianchiidtot} over the compact space $M$ leads to the well-known $D5$-brane tadpole cancellation condition,
 \eq{
  \label{tadK3}
   \int_{M} \sum_i Q_i\, \delta^{(4)}(\Delta_{6,i})+\ldots = 24\, \alpha_4(M)\, .
 }
Here the dots indicate additional  contributions
from magnetised $D9$-branes, that are not yet captured by the K-theory
groups we consider.

This relation makes it evident that
just a geometric $[M]\ne 0$ is not a topologically consistent
compactification of the type I string. Similarly, $D5$-branes
on a flat toroidal space are obstructed, too. 
Indeed, the left hand side of \eqref{tadK3}
can be  considered  as the $KO_4(pt)$  contribution and the right hand side as
the $\Omega_4^{\rm Spin}(pt)$  contribution. Each alone is
inconsistent and therefore gives a $\mathbb Z$-valued obstruction,
but by combining them one can find configurations of vanishing $C_6$-form
charge.  Note that \eqref{tadK3} means that only for multiples of $24$ $D5$-branes the
charge can be cancelled.

More precisely, this gauging can be encoded in  a map
\eq{
  \label{hsv}
   0=\Omega_4^{{\rm Spin}+D5;\,{\rm U}(1)_{(2)}}(pt)
  \longrightarrow \Omega_4^{{\rm Spin}}(pt)\oplus KO_4(pt)=\mathbb Z\oplus \mathbb Z\, ,
}  
 where the U$(1)_{(2)}$ factor indicates the two-form gauge symmetry. This map
is defined by forgetting the $C_2$-form gauge field (while
keeping the $D5$-branes). Thus, one can consider the
K-theory group as the cobordism class of the defect brane, i.e.
$\Omega_4^{D5}(pt):=KO_4(pt)$.
We stress that in the definition of the cobordism group on the left we have to include the
information about the charged objects, i.e.~we have to impose the
Bianchi identity \eqref{Bianchiidtot} and hence a topological trivialisation of
its right hand side.

Let us now comment on  the charge normalization
  $k=24$. We present an argument why this normalization
  in the tadpole condition is related to a certain string cobordism group. 
Notice that the definition of the cobordism groups $\Omega_4^{{\rm Spin}+D5;\,{\rm U}(1)_{(2)}}(pt)$ is
analogous to that of the more familiar so-called string-cobordism
groups $\Omega^{\rm String}_n(pt)$. Recall from section
\ref{subsec_string} that  the latter come with a
2-form $B$ and a trivialisation of the class $p_1/2=
24\hat\alpha_4$, which in \eqref{Bianchiidtot} is generalized to include 
also the $D5$-branes, leading to a trivialisation of the whole class
on the right hand side of the Bianchi identity. Thus, by ignoring the
$D5$-branes, for $k=24 m$  ($m$ integer) we can define the
generalized string cobordism groups 
\eq{
             \Omega_n^{{\rm Spin};\,{\rm U}(1)_{(2)},k}(pt)=\Omega^{{\rm String},k}_n(pt)\,,
             \label{SpinU12}
           }
with a trivialization of the class ${k\over 48} \, p_1 = \frac{m}{2}\,p_1 \in H^4(M,\mathbb{Z})$.

Now, due to the factor $k$ in \eqref{Bianchiidtot}, not the full $\mathbb Z\oplus \mathbb Z$
global symmetry on the right hand side of \eqref{hsv} is co-killed,
rather a discrete  $\mathbb Z_{k}\subset KO_4(pt)$
symmetry survives. This corresponds  to  a $D5$-brane  stack with less
than $k$ branes, so that it cannot be compensated by changing $\hat\alpha(M)$.
The $D5$-branes happen 
to be precisely the right defects which can break the global symmetry
of the  generalized string cobordism group\footnote{Here, we are using the fact that to break a
  global symmetry in $\Omega^{{\rm Spin};\, {\rm U(1)}_{(n-2)}}_{n-1}$ we need
  defects (branes) of the dimension related to $KO_n(pt)$.}, if 
$\Omega_3^{{\rm String},k}(pt)=\mathbb Z_{k}$.
Hence, this reasoning has led us to a close relation between the
relative normalization factor $k$ in the tadpole constraint and
a certain generalized string cobordism group. The
normalization $k=24$ realised in string theory can however not be deducted from this argument.

Notice also that the tadpole condition \eqref{tadK3} is in principle valid within the whole
cobordism group  and thus it can be interpreted as an off-shell
condition. In other words, the cobordism language allowed us to go
beyond a strictly background-dependent analysis. This observation is
in fact general and valid also for the other tadpole cancellations
that we are going to recover in the following.

To summarise, we realise  that after gauging, the (open string) K-theory groups and the
(closed string) geometric cobordism groups are really two sides of the
same coin. In fact, the K-theory groups are
the mathematically precise definition of the cobordism groups of
D-brane  defects. We will present more insights below, when discussing the Spin$^c$-cobordisms.

\subsubsection*{Gauging  discrete global symmetries}

What about the $\mathbb Z_2$-valued classes $\Omega_1^{{\rm Spin}}/KO_1(pt)$ and
$\Omega_2^{{\rm Spin}}/KO_2(pt)$, where we start with a $\mathbb
Z_2\oplus \mathbb Z_2$ global symmetry in each case?
While in  type IIB theory the global symmetries related to
$\Omega_{1\,(2)}^{{\rm Spin}}$ were argued to be
broken  with (wrapped) $O7$-planes playing the role of the
defects\cite{McNamara:2019rup}, the Hopkins--Hovey isomorphism
suggests that at least the K-theory induced global symmetries are gauged.

Thus, following the same logic as in the previous example, e.g. for
the case $\Omega_2^{{\rm Spin}}/KO_2(pt)$
we will have a $\mathbb Z_2$-valued charge neutrality condition
 \eq{
  \label{tadS1xS1}
   \int_{M} \sum_i Q_i\, \delta^{(2)}(\Delta_{8,i})
   = k\, \alpha_2(M)\qquad
   {\rm mod}\ 2\,.
 }

Here, on the left hand side are the charges of the non-BPS
$\widehat{D7}$-branes, while on the right hand side the contribution from the cobordism group.
Note that the map $\alpha_2$ is identical to the
Arf-invariant\cite{Wan:2018bns}
\eq{
  \alpha_2 ([M])=\text{Arf} (M)\,,
}  
which for the generator $M=S^1_p\times S^1_p$ is equal to one.

As before, we do not know a priori what the value $k$ of the relative
normalization is. If it is even, then the right hand side of \eqref{tadS1xS1}
decouples from this relation and the K-theory charge by itself has
to be even. In other words, the global symmetry of $KO_2(pt)=\mathbb
Z_2$ is gauged and the one of $\Omega_2^{{\rm Spin}}=\mathbb Z_2$
remains unbroken at this stage and thus needs to be broken
by introducing appropriate defects. In such a case, the wrapped $O7$-planes
of type IIB cannot do the job, as they are projected out in type I
theory. Therefore, as also suggested for the S-dual heterotic string
in \cite{McNamara:2019rup}, there must exist  new so far unknown defects.

The second possibility is that the normalization $k$ is odd. In this
case, placing for instance a single non-BPS
$\widehat{D7}$-brane on the background $M=S^1_p\times S^1_p$ would give a
neutrally charged configuration and thus it would be allowed.  Then, one linear combination of the
$\mathbb Z_2$-charges is gauged and the orthogonal one is broken. They both 
lie in the cokernel of the map
\eq{
0=\Omega_2^{{\rm Spin}+\widehat{D7};\,\mathbb Z_2}(pt) \longrightarrow
\Omega_2^{\rm Spin}\oplus KO_2(pt)=
  \mathbb Z_2\oplus \mathbb Z_2\,.
}
However, since just a torus $S^1_p\times
S^1_p$ is expected to be a valid background of type I string theory
without placing a non-BPS
$\widehat{D7}$-brane on it, this second possibility does not seem to
be realized. Thus, we conclude that $k$ is (likely) even.

The story will repeat itself for the other $\mathbb Z_2$ K-theory classes: 
$KO_1(pt)$, $KO_9(pt)$ and $KO_{10}(pt)$, which are related to the non-BPS
$\widehat{D8}$, $\widehat{D0}$ and $\widehat{D(-1)}$-branes respectively.

\subsection{Spin$^c$-structure in F-theory: $\Omega_{2+4k}^{{\rm Spin}^c}(pt)$}
\label{subsec_spinc}

The non-vanishing classes $K_{n}(pt)=\mathbb Z$ ($n$ even)  correspond
to the usual $Dp$-branes of type IIB with $p=d-n-1$.
The associated global $p$-form symmetry is gauged,
as there are appropriate RR $(p+1)$-form gauge fields $C_{p+1}$.
Thus, the story is expected to be analogous to the $KO_{4}(pt)=\mathbb
Z$ example from the previous section.
In particular, appropriate combinations of the global symmetries associated to the groups
$K_{n}(pt)$ and $\Omega^{{\rm Spin}^c}_{n}(pt)$ should be gauged.

As we have seen from the ABS-orientation, the Todd class provides the map $\alpha_n^c$ between cobordism and K-theory. Therefore, a natural candidate for the global
  current is
  \eq{
    \label{currentbord}
    \star J_{d-n}=\alpha^c([M])={\rm td}_n(M),
 }
 which indeed leads to a conserved current satisfying  $d\, \star\, J_{d-n}=0$.
 For all generators $\Sigma_n$ from table \ref{table:bordclassspinc}, the charge is 
  \eq{   
    \int_{\Sigma_n}  \star J_{d-n}=\int_{\Sigma_n} {\rm td}_n(\Sigma_n) =1\,.
  }
  The picture that will emerge from the following analysis is that 
  in the case where we are dealing only with backreacted $(d-n)$-dimensional
  localised sources, which in these set-ups are
  e.g.~$O_{d-n-1}$-planes, the  current
  will be the Todd genus. 
  For set-ups with different types of sources, 
  like the $n=6$ case below, the gauged current turns
  out to be a linear combination of the cobordism invariants,
  in particular it includes also invariants in the kernel of the ABS-orientation.
  Let us first consider the two cases $n=2$ and $n=6$ in more detail,
  which turn out to be related to F-theory.
  
\subsubsection*{Gauging the global symmetry of  $\Omega^{{\rm Spin}^c}_{2}(pt)$} 
  
 Recall that for $n=2$ the K-theory classes describe $D7$-branes and the generator
 of $\Omega_2^{{\rm Spin}^c}(pt)$ is $[\mathbb P^1]$.
After gauging and combining charged K-theory and cobordism objects, we
arrive at a magnetic $C_0$-form tadpole cancellation condition which reads
\eq{
  \label{ftheorytad}
   \int_{\mathbb P^1}  \sum_i { Q}_i \,\delta^{(2)}(\Delta_{8,i})
   =24\, \alpha^c_2([\mathbb P^1])=12\, \int_{\mathbb P^1} c_1(\mathbb P^1)\,.
 }
The normalisation factor $k=24$ on the right hand side has been fixed such that this
equation becomes precisely the well-known relation from F-theory
\cite{Vafa:1996xn} on a $K3$ surface.  In this context, the $\Delta_{8,i}$ are the divisors over which the elliptic
fiber becomes singular of degree ${ Q}_i$ and the $\mathbb P^1$ is the result of the
backreaction of the $(p,q)$ 7-branes. Hence, for the manifold  $\mathbb P^1$ itself
we really deal with an on-shell configuration.
One might
wonder whether there is a similar tadpole constraint for the NS-NS
eight-form, which is the magnetic dual of the dilaton. In F-theory
this constraint seems to be always satisfied among the $(p,q)$
7-branes so that it does not show up as a global symmetry related to
a non-trivial cobordism class.

In agreement with the previous obstruction,
both a single $D7$-brane in K-theory and a pure geometric $[\mathbb P^1]$ in cobordism are
inconsistent.
However, a combination of the two ingredients can cancel the tadpole,
i.e.~it can lead to a configuration with vanishing $C_0$-form charge.
This gauging  is expected to correspond to a map
\eq{
  \label{werder}
  0=\Omega_2^{{\rm Spin}^c+D7;\, {\rm U}(1)_{(0)}}(pt) 
  \longrightarrow \Omega_2^{{\rm Spin}^c}(pt)\oplus K_2(pt)=\mathbb Z\oplus \mathbb Z,
}  
where the U$(1)_{(0)}$ is the RR eight-form gauge symmetry whose total
Bianchi identity and implied trivialisation has to be imposed in the definition of the full
cobordism group $\Omega_2^{{\rm Spin}^c+D7;\, {\rm U}(1)_{(0)}}(pt)$.
Similarly to the $K3$ example from section \ref{subsec_spintad}, a discrete
$\mathbb Z_{24}\subset  K_2(pt)$ symmetry is not co-killed. We conjecture that the
relative normalisation $k=24$ will show up in
\eq{
  \Omega_1^{{\rm Spin}^c;\, {\rm U}(1)_{(0)},k}(pt)=\mathbb Z_k\,.
}
The actual determination of such cobordism groups is left for future studies.

Moreover,  on the purely mathematical level, 
the  ABS homomorphism \eqref{abshomospinc}  seems to indicate
that $[\mathbb P^1]\in  \Omega_2^{{\rm Spin}^c}$ carries the same
magnetic $C_0$-form charge as a single $D7$-brane, which is the
generator of $\widetilde{K}_2(S^2)$. However, we have seen  that
one needs $k=24$ $D7$-branes to get a $\mathbb P^1$ upon backreaction.
This might suggest that the physically
relevant generalisation of the ABS map is rather 
\eq{
  \alpha^c_2([M])= 24\, {\rm Td}(M)\in 24\,\mathbb Z\,,
}
which is still a group homomorphism, but ceases to be surjective.
As alluded to earlier,  the intuitive arguments from section \ref{sec_KCisom}
for the existence of a map
between cobordism and K-theory are meant up to such relative
normalisation factor.

\subsubsection*{Gauging the global symmetry of  $\Omega^{{\rm
      Spin}^c}_{6}(pt)$}

For $n=6$ we are dealing with $D3$-branes classified by
$K_6(pt)=\mathbb Z$ and the cobordism group $\Omega_6^{{\rm
    Spin}^c}(pt)=\mathbb Z\oplus \mathbb Z$. Recall that the first
$\mathbb Z$ is the one whose charge is measured by the Todd class, the
charge
of the second is instead determined by $c_1^3/2$. The diagonal generators of the
two were denoted as  $[\mathcal H_1]$ and $[\mathcal H_2]$, where the
second one generates also the  kernel of $\alpha^c_6$.
Gauging the global three-form symmetry related to the Todd class, one expects that there exists a
$C_4$-form tadpole cancellation condition such that
\eq{
  \label{ftheorytadd3}
    \int_{\mathcal H_1}  \sum_i Q_i\,\delta^{(6)}(\Delta_{4,i}) =\gamma \, \alpha^c_6([\mathcal H_1])=
    {\gamma\over 24} \int_{\mathcal H_1}  c_2(\mathcal H_1)\, c_1(\mathcal H_1)\,,
}
where the normalisation $\gamma$ needs to be determined.
Recall that a connected representative for
  $[\mathcal{H}_1]$ is the threefold $[dP_9\times \mathbb P^1]$,
  which can be thought of as the downstairs geometry of the orientifold
  $K_3\times T^2/\Omega (\sigma, I_2)$. Here, as in
  \cite{Dabholkar:1996zi,Gimon:1996ay}, $\sigma$ is a holomorphic
  involution of $K3$ and $I_2$ the reflection of the two coordinates of
  $T^2$. Note that this orientifold only has $O3$-planes and no $O7$-planes.

To proceed, recall the $D3$-brane
tadpole cancellation condition \cite{Sethi:1996es} for F-theory compactified on a smooth
elliptically fibered Calabi-Yau fourfold $Y$ over a base $B$
\eq{
  \label{ftadfour}
     \int_{B} \sum_i  Q_i\,\delta^{(6)}(\Delta_{4,i})+\ldots ={\chi(Y)\over 24}=\int_{B} \left({1\over 2}\, c_2(B)\, c_1(B) +15\, c_1^3(B)\right)\,.
}
Since the second term vanishes for the generator $\mathcal H_1$, the
two conditions \eqref{ftheorytadd3} and \eqref{ftadfour} coincide for
$\gamma=12$.
Note that  in the Sen limit \cite{Sen:1997gv} the  right hand side also includes
contributions from geometrized $O7$-planes and not only $O3$-planes.

Remarkably, the general
tadpole condition \eqref{ftadfour} contains precisely the two
cobordism invariants of $\Omega_6^{{\rm Spin}^c}(pt)$, in particular $c_3(B)$ is absent. One might wonder what the role of the second charge
$\mathbb Z$ is. Since both $\mathbb Z$-charges appear in
\eqref{ftadfour}, it is a linear combination of the  two that is gauged and
couples to the  $C_4$-form. The current then is  
\eq{
    \label{currentbordb}
    \star J_{4}\sim 12\,{\rm td}_6(B) +30\left( {c_1^3(B)\over 2}\right) \,.
} 
This gauging  is expected to correspond to a map
\eq{
  \label{hannover96}
  0=\Omega_6^{{\rm Spin}^c;\, {\rm U}(1)_{(4)}}(pt) 
  \longrightarrow \Omega_6^{{\rm Spin}^c}(pt)=\mathbb Z\oplus \mathbb Z,
}
where the U$(1)_{(4)}$ is the four-form gauge symmetry, whose Bianchi
identity and the implied trivialisation are again imposed. Then, both global symmetries
from  $\Omega_6^{{\rm Spin}^c}(pt)$ lie in the
cokernel and are therefore co-killed.

We think that intuitively the appearance of two initial global symmetries is related to the fact
that both pure $O3$-planes as well as wrapped  $O7$-planes
contribute to the right hand side of the tadpole equation.
This means that prior
to gauging, the bulk duals  of pure $O3$-planes  and wrapped
$O7$-branes  are not cobordant in $\Omega_6^{{\rm Spin}^c}(pt)$ and
give rise to two independent global symmetries.

\subsection{Spin$^c$-structure in F-theory: $\Omega_{4k}^{{\rm
      Spin}^c}(pt)$}

We discuss now the remaining Spin$^c$-cobordism groups. We
will be led to tadpole constraints similar to those of the previous sections, 
whose form is however not as clearly established.

\subsubsection*{Gauging the global symmetry of  $\Omega^{{\rm
      Spin}^c}_{0}(pt)$}

For $n=0$, the K-theory group classifies space-time filling $D9$-branes
so that the appearing global symmetry will be gauged via $C_{10}$.
The related cobordism group $\Omega^{{\rm Spin}^c}_{0}(pt)=\mathbb Z$
should then provide the geometric contribution to the $C_{10}$ tadpole
cancellation condition. In this case, the only natural candidate for
the generator  ${\rm pt}^+$ is the  space-time filling $O9$-plane.
Thus, there exist essentially two solutions of the resulting tadpole
constraint, namely just the trivial 10D type IIB theory or the type I superstring.

\subsubsection*{Gauging the global symmetry of  $\Omega^{{\rm
      Spin}^c}_{4}(pt)$}

For $n=4$, the cobordism group $\Omega^{{\rm Spin}^c}_{4}(pt)=\mathbb Z\oplus \mathbb Z$
suggests that the story can be similar to the $n=6$ case above. 
However, there are few subtleties which we discuss below and which
can lead in fact to different possibilities.
  
For $[K3]=2[{\mathcal G}_1]$ one gets the
same structure as for $\Omega^{{\rm Spin}}_{4}(pt)$ described in
section \ref{subsec_spintad},
in particular a type I background with $D9$ and $D5$-branes.
Of course, here we are assuming that one cancels  the $C_{10}$-tadpole
in a non-trivial way by an $O9$-plane and 32 $D9$-branes.
However, the two generators  of $\Omega^{{\rm Spin}^c}_{4}(pt)$ are
not Calabi-Yau so that  following our logic, there should exist a $C_2$ tadpole condition like
\eq{
  \label{ftad5}
     \int_{B} \sum_i  Q_i\,\delta^{(4)}(\Delta_6^i)+\ldots =\int_{B}
     \left( 12\,  {\rm td}_4(B) -{3\over 2} \, c_1^2(B)\right)\,.
}
The normalisation factors $12$ and $-3/2$  have been fixed by the K3 case mentioned
above and  by evaluating  the term $p_1\sim{\rm tr}(R\wedge R)$
(Pontryagin class) arising from
the Chern-Simons term in the effective action of the $D9$-branes and the
$O9$-plane. Note that for $c_1(B)$ not even, there will be another
half-integer contribution on  the left hand side from the required
line bundle ${\cal L}^2=c_1(B)$ supported on the $D9$-brane wrapping the
Spin$^c$-manifold $B$.
Similarly to the $n=6$ case, one  could consider  one
$\mathbb Z$ factor as the geometric counterpart of such wrapped
$O9$-branes and  the other one as that of genuine $O5$-planes.

Note that in contrast to $K3$ the generators of $\Omega^{{\rm
    Spin}^c}_{4}(pt)$, namely the manifolds $B=\mathbb P^1\times \mathbb P^1$ and
$B=\mathbb P^2$,  are off-shell configurations of type I.
Could there be a different interpretation of the constraint
\eqref{ftad5} that is not using the type I superstring and where the
 manifold is a generator of  $\Omega^{{\rm
    Spin}^c}_{4}(pt)$ and on-shell?
Recall that there exist  also supersymmetric type IIB orientifolds of $K3$
with  the $\mathbb Z_2$ projection $\Omega \sigma$,
where $\sigma$ is a holomorphic involution
\cite{Dabholkar:1996zi,Gimon:1996ay}.
This  leads to $O5$-planes
at the fixed points of $\sigma$. Downstairs the manifold is
$[K3/2]=[dP_9]$ so that it is suggestive to  get a geometrized $D5$-brane
tadpole condition of the form
\eq{
  \label{ftad5c}
     \int_{dP_9} \sum_i  Q_i\,\delta^{(4)}(\Delta_6^i) =k\int_{dP_9}   {\rm td}_4(dP_9) \,.
   }
Here, we used $\int {\rm td}_4(dP_9)=1$ and $\int c_1^2(dP_9)=0$,
leading to the identification in cobordism  $[dP_9]=[{\cal G}_1]$.
Note that for such a geometrized  $O5$-plane, the current is indeed given by
the ABS isomorphism, i.e. the Todd class.

An interesting possibility is that 
\eqref{ftad5c} is analogous to the F-theory-like tadpole \eqref{ftheorytad}.
Without the $O9$-planes, there are also $N\!S5$-branes present in type IIB.
Such branes   backreact on
the geometry, so that there might exist globally consistent
backreacted solutions of  in general $(p,q)$ 5-branes.\footnote{Similar
  geometrized
  configurations have been discussed in four dimensions in
  \cite{Martucci:2012jk,Braun:2013yla,Candelas:2014jma}.}
Like in F-theory, there could then exist special limits where these become
the aforementioned  perturbative $\Omega \sigma$
orientifolds of type IIB on $K3$. Comparing to the concrete toroidal
orbifold example from \cite{Dabholkar:1996zi}, such a picture would suggest $k=24$.
Of course, without a concrete construction all this is fairly speculative.

Let us also mention that on-shell backgrounds for  $B=\mathbb P^1\times \mathbb P^1$ or
$B=\mathbb P^2$ are known to appear as bases for F-theory
compactifications on elliptically fibered Calabi-Yau threefolds.
Usually,
there is no non-trivial $C_2$-form tadpole condition.
However, it is known that there exist base geometries
that via F-theory-heterotic
duality correspond to the presence of $N\!S5$-branes on the heterotic side
\cite{Morrison:1996na,Morrison:1996pp,Aspinwall:1997ye}.
With each $N\!S5$-brane comes a tensor multiplet, which in the F-theory
dual increases the number of 2-cycles in
the base, i.e.  $n_T=h^{1,1}(B)-1$, $n_T$ being the number of tensor multiplets.

On the heterotic side,
the Hirzeburch surfaces $\mathbb F_n$ correspond
 to smooth configurations with instanton numbers
 $(12+n)$ and $(12-n)$ in the two $E_8$ factors and no 5-branes.
For all $\mathbb F_n$ one finds
$\int {\rm td}_4=1$ and $\int c_1^2=8$ so that they are all (at least
 for $n$ even) cobordant\footnote{Actually, for $n$ odd, $\mathbb F_n$ is not Spin
  and one has an additional non-trivial line bundle on the Spin$^c$-manifold.}
to $\mathbb F_0=\mathbb P^1\times \mathbb P^1$.
Consistently, the right hand side of the tadpole relation \eqref{ftad5}
vanishes.
Thus, six-dimensional F-theory-heterotic duality suggests
that there might exist a relation like
\eq{
  \label{ftad5b}
      N_{NS5}\big\vert_{\rm het} =n_T-1\overset{?}{=}\int_{B}
     \left( 8\,  {\rm td}_4(B) -\, c_1^2(B)\right)\,
   }
   for the number of $N\!S5$-branes  on the heterotic
   side.\footnote{Coming back to the discussion around eq.\eqref{ftad5c},
     for $B=dP_9$ we obtain $n_T=9$ which is indeed the
     result for the $\Omega\sigma$ orientifold of \cite{Dabholkar:1996zi}.}

 The apparent controversy of this  example demonstrates that starting with incomplete
 cobordism structures has its limitations, but still has the potential
 to uncover new aspects  of string theory.

\subsubsection*{Comment on gauging the global symmetry of  $\Omega^{{\rm
      Spin}^c}_{8}(pt)$}

For $n=8$ we know that $\Omega_8^{{\rm Spin}^c}(pt)=\mathbb Z^4$, but we
have fixed neither the four cobordism invariants nor the
generators of the cobordism group. What we can say is that eventually
there should be a $C_6$ tadpole cancellation condition reflecting 
a  map
\eq{
  0=\Omega_8^{{\rm Spin}^c;\,{\rm U}(1)_{(6)}}(pt)
  \longrightarrow \Omega_8^{{\rm Spin}^c}(pt)=\mathbb Z^4\,.
}
Again, the four global symmetries
of $\Omega_8^{{\rm Spin}^c}(pt)$ could be  bulk counterparts 
of wrapped  $O9$, two orthogonally wrapped   $O5$ and localised $O1$-planes.

\section{Conclusions}

We have provided evidence that there exists
a mathematical manifestation of open-closed string duality already at the purely
topological level. Concretely, the relation is between K-theory
and certain cobordism classes, as it is described by the
generalisations of the Conner-Floyd theorem to Spin and Spin$^c$-manifolds.

Whereas the description of stable $D$-branes in terms of K-theory classes
is well established, the recent proposal for the relevance of
cobordism classes for consistent string backgrounds is rather
on the level of a conjecture. We think that the interpretation
of the Hopkins--Hovey theorem presented in this work
gives more credence to
the cobordism conjecture. In fact, in view of open-closed string duality
one is automatically driven to the relevance
of cobordism for closed string backgrounds.
One could say that it encodes topological quantum
gravity information on the consistent
coupling of the gauge sector  to gravity.
By including K-theory classes, we have extended the original
work of McNamara--Vafa \cite{McNamara:2019rup} by  giving
a more precise mathematical meaning to the cobordism
class related to defects (D-branes).

Moreover, we have exploited the consequences of this  correspondence
and discussed in a bottom-up approach the fate of global symmetries
arising in the various setups. We suggested that the physical
significance of the Hopkins--Hovey isomorphisms is that
the directly involved combinations of global symmetries are gauged in the respective
theories. As a consequence, orthogonal combinations of global symmetries were co-killed
and hence broken.
We found an intricate relation to some well-known tadpole cancellation conditions, where the bulk
contributions were linear combinations of the cobordism invariants.

While it  deserves further investigation,
the six-dimensional example $\Omega_4^{{\rm Spin}^c}(pt)$
 might be considered as evidence that this approach has the potential to
uncover also new aspects of string theory, even for the case of
removing the global symmetries via gauging.

The examples made it evident
that the intermediate appearance
of extra global symmetries is a consequence of having ignored
additional ingredients present in the final cobordism group
$\Omega_n^{\rm QG}$. Hence, it is a  relict of trying to narrow down
the QG-structure following  a bottom-up approach.
Clearly, it would be very
interesting to develop means to compute
such refined gauge cobordism classes from first principles
(see also \cite{Andriot:2022mri}).

One could  ask whether at least part of the  story can be
extended to more general setups. One can certainly  consider $D$-branes
in backgrounds with non-trivial $H$-flux, which are classified
by twisted K-theory classes $K_H(X)$. Hence, one might expect a notion
of $H$-twisted cobordism classes, as well.
There have been proposals for
cobordism groups relevant for the heterotic string ($\Omega^{\rm
  string}_n$) or even M-theory ($\Omega^{{\rm pin}^+}_n$, but see also the recent proposal \cite{Sati:2021uhj,Sati:2019tqq}, similar in spirit to our analysis). 
It would be interesting to generalize our analysis to these theories as well.


\noindent
\paragraph{Acknowledgments:}
We would like to thank Markus Dierigl, I\~{n}aki Garc\'ia-Etxebarria, Christian Knei\ss l, Andriana Makridou, Miguel Montero and Jacob
McNamara for discussions. We are also grateful to  Markus Dierigl and Andriana Makridou for comments on the draft. N.C.~would like to thank David Andriot and Nils Carqueville for discussions and collaboration on related topics. The work of N.C.~is supported by the Alexander-von-Humboldt foundation.

\vspace{0.4cm}
\clearpage

\appendix

\section{On the cobordisms invariants of $\Omega_6^{{\rm Spin}^c}(pt)$}
\label{app_todd6}

In \cite{Wan:2018bns}, the cobordism invariants of $\Omega_6^{{\rm Spin}^c}(pt)$
are listed as
\eq{
  c_1 {\sigma\over 16}\,,\qquad\quad {c_1^3\over 2}
}
with $ c_1 {\sigma\over 16}:={\sigma\over 16}({\rm PD}(c_1))$, where
${\rm PD}(c_1)$ is the submanifold $D$ of the Spin$^c$ six-manifold $B$ which
represents the Poincar\'e dual of $c_1(T_B)$. Below, we show that these invariants coincide with those employed in the main text.

Due to the Hirzebruch signature theorem, we can write
\eq{
       c_1 {\sigma\over 16}={1\over 48}\int_D p_1(T_D) = {1\over
         48}\int_B p_1(T_D)\wedge c_1(T_B),
     }
where  the first Pontryagin class can be expressed in terms of Chern
classes as $p_1=-2c_2 + c_1^2$.     
Now one invokes the normal sequence
\eq{
      0\longrightarrow T_D \longrightarrow T_B\vert_D \longrightarrow
      N \longrightarrow 0
}      
with the normal line bundle $N$ featuring $c_1(N)=c_1(T_B)$.
Using  that for the total Chern classes one has $c(T_D)=c(T_B\vert_D)/c(N)$, we
can determine $c_1(T_D)=0$ and $c_2(T_D)=c_2(T_B)$, i.e.~we find that the
hypersurface is CY. Putting everything together, we arrive at
\eq{
       c_1 {\sigma\over 16}= {1\over
         48}\int_B p_1(T_D)\wedge c_1(T_B)=-{1\over 24}\int_B
       c_2(T_B)\wedge c_1(T_B)=-{\rm Td}(B)\,.
}
This shows that, consistently with the ABS orientation, the Todd class
is  a cobordism invariant of $\Omega_6^{{\rm Spin}^c}(pt)$.
       
\section{Extension of scalars}
\label{app_eos}

Consider a ring $R$, a right R-module $M$ and a left R-module $N$. Their tensor product over $R$ is denoted as $M \otimes_R N$ and it is obtained by associating to each pair $(m,n)$ a tensor $m\otimes_R n$. This is a ring and in fact a module over $R$, in particular
\begin{equation}
r \cdot (m \otimes n) = (r\cdot m) \otimes n,\qquad  (m \otimes n) \cdot r = m \otimes (n\cdot r).
\end{equation}
We want to generalise the above to the case in which we have modules over different rings. Consider another ring $S$ together with a ring homomorphism $f: R \to S$. Let $M$ be an R-module. Being a ring, $S$ is a module over itself but also an $R$-module via the homomorphism $f$. Then, we can construct
\begin{equation}
M_S = M \otimes_R S.
\end{equation}
This is called extension of scalars and it is a module over $S$
constructed out of the R-module $M$. In this sense, it can be thought
as a change of ``ring basis'' from $R$ to $S$.
Due to the action of $f$, there is a natural equivalence relation
\begin{equation}
(m  \cdot r) \otimes s \sim m \otimes (f(r) \cdot s),
\end{equation}
which allows us to define $M_S$ as the $S$-module
\begin{equation}
M \otimes_R S = \{ m\otimes s, \,\, \text{s.t.}\,\, \forall r \in R,\, (m  \cdot r) \otimes s = m \otimes (f(r) \cdot s)\}\,.
\end{equation}

\section{Proof of the isomorphism \eqref{RMisom}}
\label{app_proof}

In this appendix we present the proof of the standard isomorphism
$R \otimes_R M \cong M$,
which is valid for any ring $R$ (which is also a module over itself)
and  any $R$-module $M$.
We define a bilinear map $h: R \times M \to M$, such that $(r,m) \to r \cdot m$. By the universal property of the tensor product, this induces a linear map $f: R\otimes_R M \to M$, such that $r\otimes m \to r \cdot m$. The inverse linear map $g: M \to R \otimes_R M$ sends $m \to 1 \otimes m$. The isomorphism follows since $ g \circ  f = \mathbb{I}_M$ and $f \circ g = \mathbb{I}_{R\otimes_R M}$.



\clearpage

\bibliography{references}  
\bibliographystyle{utphys}


\end{document}